\documentclass[useAMS,usenatbib]{mn2e}
\usepackage{epsfig}
\begin{document}

\title[SN dust microphysics] {On the effects of microphysical grain 
properties on the yields of carbonaceous dust from type II SNe} 
\author[Fallest et al.]
{David W. Fallest$^{1}$\thanks{E-mail: dwfalles@unity.ncsu.edu}, 
Takaya Nozawa$^{2}$, Ken'ichi Nomoto$^{2}$, 
Hideyuki Umeda$^{3}$, \newauthor Keiichi Maeda$^{2}$, Takashi Kozasa$^{4}$ and Davide Lazzati$^{1}$\\
$^{1}$ Department of Physics, NC State University, 2401 Stinson 
Dr., Raleigh, NC 27695-8202 \\
$^{2}$ Institute for the Physics and Mathematics of the Universe, 
University of Tokyo, Kashiwa 277-8583, Japan \\
$^{3}$ Department of Astronomy, School of Science, University of 
Tokyo, Tokyo 113-0033, Japan \\
$^{4}$ Department of Cosmosciences, Graduate School of Science, 
Hokkaido University, Sapporo 060-0810, Japan}

\maketitle

\begin{abstract}
  We study the role of the unknown microphysical properties of
  carbonaceous dust particles in determining the amount and size
  distribution of carbonaceous dust condensed in type II supernova
  explosions. We parametrize the microphysical properties in terms of
  the shape factor of the grain and the sticking coefficient of
  gas-phase carbon atoms onto the grain surfaces. We find that the
  amount of dust formed is fairly independent of these properties,
  within the parameter range considered, though limited by the
  available amount of carbon atoms not locked in CO molecules.
  However, we find that the condensation times and size distributions
  of dust grains depend sensitively on the microphysical parameters,
  with the mass distributions being weighted toward larger effective
  radii for conditions considering grains with higher sticking
  coefficients and/or more aspherical shapes.  We discuss that this
  leads to important consequences on the predicted extinction law of
  SN dust and on the survival rate of the formed grains as they pass
  through the reverse shock of the SN. We conclude that a more
  detailed understanding of the dust formation process and of the
  microphysical properties of each dust species needs to be achieved
  before robust prediction on the SN dust yields can be performed.
\end{abstract}


\section{Introduction}

Interstellar dust, once considered to be little more than a nuisance
to astronomical observations, is one of the most interesting areas of
astrophysical research today \citep{LG03}. One aspect of particular
interest is the origin of dust at high redshift ($z>5$). Possible
sources of interstellar dust that have been considered include
outflows from asymptotic giant branch (AGB) stars \citep{ME03, ZG08,
  VS09}, Wolf-Rayet systems \citep{IC10}, quasars \citep{EM02}, and
supernova explosions \citep{KH89, KH91, TF01, NK03, BS07, KN09,
  IC10}. How much dust can be attributed to each of these possible
sources at such high redshift remains unclear. Supernovae (SNe) are
considered by some to be likely contributors of much of the dust in
the early Universe because their progenitors are quite massive and
consequently have short lifetimes \citep{BC02, ME03, MS04,
  DG07}. However, such early SNe have not been observed, making
determinations of their dust contributions difficult. Instead we need
to consider dust yield predictions of more recent SNe, which have been
observed, and then extrapolate the dust yields to earlier SNe. The
theoretical predictions of dust yields for recent supernovae
\citep{KH91, TF01, NK08}, however, are too large compared to
observations, with discrepancies that can be as high as 3 to 4 orders
of magnitude in some cases \citep{HW70, LD89, WR93, AE03, WM07, RK09}. Observations of young supernova remnants (SNR), however, have confirmed dust masses one to two orders of magnitude greater than SNe. Some of these SNR have observed dust masses of 0.02--0.054 M$_{\odot}$ by {\it Spitzer} \citep{RK08}, 0.06 M$_{\odot}$ by {\it AKARI} \citep{SA10}, and 0.075 M$_{\odot}$ by {\it Hershel} \citep{BK10}, in Cassiopeia A, and 0.04--0.1 M$_{\odot}$ by {\it Spitzer} in the pulsar wind nebula G54.1+0.3 \citep{TS10}. Additionally, \citet{NK10} have demonstrated that the observed spectral energy distribution of Cassiopeia A can be well reproduced by the calculations of dust formation in Type IIb SNR and the mass in the SNR is 0.07 M$_{\odot}$. 
 
Reconciling the dust yield prediction for local SNe can be done via
one of two channels: improving our understanding of the dust formation
process, leading to a substantially decreased prediction, or revising
the observational constraints to account for a higher dust yield than
so far implied. It is possible that some amount of dust has avoided
observation. Dust particles absorb light and re-emit the energy at
infrared wavelengths. However, cold dust at temperatures
of a few tens of Kelvin could escape detection at mid-infrared
wavelengths. Additionally, areas where the dust is optically thick
could obscure some amount of dust, again allowing some dust to not be
detected. Dust clumping may also affect the estimates of dust mass
from absorption \citep{BS06}, since it is usually assumed that the
surface filling factor of dust is close to unity, while substantial
clumping could be present due to the intrinsic inhomogeneity of the
ejecta and Rayleigh-Taylor instabilities in the expanding ejecta
\citep{DW97, DL99}.

On the theoretical side, all the estimates for dust production in SNe
are based on the so-called classical theory of nucleation (e.g.,
\citet{BD35}).  It has been argued, however, that the use of classical
nucleation theory in astrophysical environments is questionable
\citep{DN85, DL08}. In addition, almost all SN dust nucleation models
thus far have considered the formation of spherical grains, and
assumed any atoms/molecules that contact the grain will adhere to the
grain; conditions that reflect maximally efficient nucleation. It is
therefore not entirely surprising that the theoretical estimates of SN
dust yields, based on upper limit of nucleation efficiency, are in
excess of those from observations. It should be noted that
\citet{BS07} have considered less than maximally efficient nucleation
by assuming the probability of atoms/molecules adhering to the grain
is less than unity, resulting in smaller dust yields.

Since dust formation is a highly non-linear phenomenon, understanding
the effects of different nucleation rates on the final dust yields is
difficult. To check the effects of different nucleation conditions we
have performed a parametric study, in which we consider carbonaceous
dust production in a SN explosion by varying the shape of the forming
grains as well as the sticking coefficient, i.e., the probability that
an incoming monomer will stick to the grain rather than bounce off and
remain in the gas phase. Our study is phenomenological and aims at
understanding which conclusions of previous nucleation studies are
robust to changing the parameters, and which may need to be
investigated more thoroughly.  A self-consistent nucleation model in
astrophysical conditions will be achieved by involving a kinetic
approach \citep{DN85, DL08, KL11} and detailed chemistry of precursor
molecules \citep{CD09, CD10}. Such a detailed approach, however, is
still under development and is not yet applicable to large scale
simulations like the one that we use here, and that have been used in
previous investigations of e.g., \citet{TF01, NK03, NK10}.

This paper is organized as follows: in Sect.~2 we detail the
nucleation theory that we adopted; in Sect.~3 we describe the
numerical code used for the computations; and in Sect.~4 we describe
our results. In Sect.~5 we finally discuss the implication and
limitations of our results and lay out future perspectives for SN dust
studies.

\section{Nucleation}
  
Nucleation is the first step of a first-order phase transition. In the
case we consider here, the phase transition is from a gas of carbon
atoms to solid clusters of amorphous carbon dust. There is a phase
equilibrium pressure where both the gas and solid phases are stable
within a given volume. The phase equilibrium pressure depends on the
temperature of the materials in the gas phase within the
volume. Nucleation of clusters of atoms/molecules of the new phase is
favoured when the gas phase is supersaturated (i.e., the pressure of
the gas is higher than the phase equilibrium pressure). The higher the
supersaturation (i.e., the ratio of the pressure to the phase
equilibrium pressure), the smaller the size of the stable clusters
that are able to form. The size of the smallest stable cluster able to
form at a given temperature and density is called the critical cluster
size. Clusters that are smaller than the critical size will tend to
evaporate, while larger clusters will tend to continue to grow. The
goal of any nucleation theory is to calculate how many critical
clusters form per unit volume per unit time.

The classical theory of nucleation considers nucleation as a
thermodynamical process in quasi-equilibrium \citep{BD35, FR66,
  DK00}. Besides the supersaturation level, the physical properties of
the nucleating material affect the size of a critical cluster.  The
classical nucleation theory assumes that all clusters share the same
properties, such as surface tension and shape, independent of their
size. The theory also assumes that those properties are equal to those
of a macroscopic sample. Moreover, the clusters are assumed to have a
uniform equilibrium temperature that is equal to the temperature of
the surrounding gas. These basic assumptions are problematic because
macroscopic thermodynamic properties are not expected to be applicable
to clusters of only a few atoms.

A different approach is provided by the kinetic theory of nucleation,
which is applicable to very small cluster sizes. The kinetic theory
relies on calculating the attachment and detachment frequencies of
monomers to a cluster \citep{DK00}. Furthermore, in contrast to the
classical nucleation theory, the kinetic theory follows the formation
of clusters smaller than the critical cluster size. In the framework
of the kinetic theory the critical cluster is the cluster whose
attachment and detachment frequencies are equal, and thus it is
stable. The downside of the kinetic theory is that attachment and
detachment frequencies for all cluster sizes need to be calculated in
order to determine the overall nucleation rate. The main aim of this
paper is to reveal the dependence of dust formation processes on the
microphysical properties of grains. In order to achieve this, we adopt
the simpler classical nucleation theory, rather than using the kinetic
theory that demands a more complicated treatment.

When the supersaturation level is $S>1$, nucleation can take place
because the free energy of the new phase is lower than that of the old
phase \citep{DK00}. The change in the free energy is due to the work
necessary to form the critical size clusters. Energy is released in
the formation of the volume of the cluster, but energy is required in
order to form the surface of the cluster \citep{DK00}. The nucleation
rate (Eq.~(\ref{eq:j1})) is given basically by two factors: the number
density of critical clusters and a kinetic factor that describes the
rate at which clusters become large enough to be stable (i.e.,
critical size or larger).

The general equation of stationary nucleation is given by
(\citet{DK00}, their equation~13.39):
\begin{equation}
J_{S} = A\exp{(-W^{*}/kT)},
\label{eq:j1}
\end{equation}
where $A$ is the kinetic factor, $W^{*}$ is the work needed to form
the critical size cluster from the gaseous state, $k$ is the Boltzmann
constant, and $T$ is the gas temperature. The kinetic factor, $A$,
is given by $A = A'\exp{(\Delta \mu/kT)}$, and $A'$ is:
 \[A' = \gamma^{*} \left( \frac{ c^{3} \sigma }{ 18 \pi^{2} m_{0} }
 \right)^{ \frac{1}{2} } \left( \frac{ p_{e} v_{0} }{ kT }
 \right)C_{0},\] 
 where $\gamma^{*}$ is the size dependent sticking coefficient, $c$ is
 the shape factor of the cluster, $\sigma$ is the surface tension,
 $m_{0}$ and $v_{0}$ are, respectively, the mass and volume of the monomer of
 nucleating material, $p_{e}$ is the phase-equilibrium pressure, and
 $C_{0}$ is the concentration of sites where nucleation can occur
 (equations 13.41 and 13.44, respectively in \citet{DK00}).  The
 concentration of gaseous monomers, $C_{1}$ is related to the
 concentration of nucleation sites by $C_{1} =
 C_{0}\exp{(-W_{1}/kT)}$, where $W_{1}$ is the work needed to form a
 cluster consisting of one monomer (equation 7.5 in \citet{DK00}). We
 consider the monomer in the gaseous state to be indistinguishable
 from the monomer in the condensed state, so that $W_{1} = 0$ and
 $C_{1} = C_{0}$. Using the supersaturation ratio, $S = C_{1}/C_{1,e}
 = p/p_{e}$ and $\Delta \mu = kT\ln{S}$, where $C_{1,e} = p_{e}/kT$ is the gaseous monomer concentration at
 the phase-equilibrium pressure and $p$ is the partial pressure of
 gaseous monomers, 
 we obtain
\[A = \gamma^*\left( \frac{c^3
     v_{0}^{2}\sigma}{18\pi^2m_{0}} \right)^\frac{1}{2} C_{1}^{2}.\]
 While the sticking coefficient may depend on the size of the cluster,
 we consider it to be constant, so that
 $\gamma^{*}=\gamma=$~constant. Finally, using the work to form a
 critical sized cluster $W^{*} =
 4c^{3}v_{0}^{2}\sigma^{3}/27\Delta\mu^{2}$ (\citet{DK00}, equation
 4.8), we find our stationary nucleation rate equation to be:
\begin{equation} 
J_{s} = \gamma\left(
     \frac{c^{3}v_{0}^{2}\sigma}{18\pi^{2}m_{0}}
   \right)^{\frac{1}{2}}C_{1}^{2}\exp{\left(
       \frac{-4c^{3}v_{0}^{2}\sigma^{3}}{27(kT)^{3}(\ln{S})^{2}}
     \right)}.
\label{eq:Js}
\end{equation}

After nucleation the clusters grow through impingement of monomers upon
the cluster. To find how much the clusters grow we begin by finding the
volume of the newly nucleated cluster. The clusters nucleate with some
critical number of monomers, $n^{*}$ (equation 4.7 in \citet{DK00}):
\begin{equation} 
n^{*} = \frac{8c^{3}v_{0}^{2}\sigma^{3}}{27\Delta \mu^{3}}. 
\label{eq:nstar} 
\end{equation} 
The volume of the critical cluster is then, $v^{*} = n^{*}v_{0}$. The
change in the cluster's volume over time depends on the sticking
coefficient $\gamma$, the surface area of the cluster
$\Sigma$, the concentration of monomers $C_{1}$, the volume of the
monomer $v_{0}$, and the average relative speed of the monomers with
respect to the cluster. In this paper we consider clusters that can
nucleate with aspherical shapes through the use of the shape factor
$c$. The shape factor is a dimensionless quantity that relates the
surface area $\Sigma$ of an object to its volume $V$ by:
$c=\Sigma/V^{2/3}$ \citep{DK00}. Thus we find the change in volume
over time to be:
\begin{equation} 
\frac{dV}{dt} = \gamma c
V^{\frac{2}{3}}C_{1}v_{0}\left( \frac{kT}{2\pi m_{0}}
\right)^{\frac{1}{2}}. 
\label{eq:dvdt} 
\end{equation}

In the same manner as \citet{NK03} we compute the depletion of the
available nucleation material through mass conservation:
\begin{equation} 
1 - \frac{C_{1}(t)}{\tilde{C}_{1}(t)} = 1-Y_{1} = \int^{t}_{t_{e}}\frac{J(t')}{\tilde{C}_{1}(t')}\frac{V(t,t')}{v_{0}}dt',
\end{equation} 
where $\tilde{C}_{1}$ is the nominal concentration of monomers -- the
concentration expected should nucleation not occur, which in an
expanding shell of volume $V_{shell}$ can be found using
\[\tilde C_{1}(t=t_{n})V_{shell}(t=t_{n}) = \tilde C_{1}(t=t_{0})V_{shell}(t=t_{0}),\]
so that the total number of gas-phase atoms remains constant, and $V(t,t')$ is the volume of a cluster
formed at time $t'$ and measured at time $t$. Rather than computing the integral on the right-hand side, we instead calculate $C_{1}(t)$ as described in step (iii) in the next section. From
Eqs. (\ref{eq:Js}) and (\ref{eq:dvdt}), we
see that $J_s$ and $dV/dt$ are simply proportional to the sticking
coefficient $\gamma$. Thus, we expect that reduced sticking coefficients
will suppress both nucleation and grain growth. The dependence of
$J_s$ on the shape factor $c$ is more complicated, since it appears in
both the kinetic factor and the exponential term. The shape factor
in the exponential term, however, will dominate the nucleation rate
equation and we expect that increased shape factors will suppress
nucleation. On the other hand, grain growth
($dV/dt$) is simply proportional to $c$, and thus an increase in the
shape factor will increase the cluster growth rate.

\section{Simulations}

We concentrate on the formation of carbonaceous grains (clusters) from
carbon atoms in the expanding material of a core-collapse
supernova. Our simulations are based on the hydrodynamic results and
elemental composition for the unmixed ejecta of a core-collapse
supernova of a $20$~M$_{\odot}$ progenitor star with metallicity $Z=0$ and an explosion energy of $10^{51}$\ ergs by \citet{UN02} (see also \citet{KN06}).

\begin{figure}
\psfig{file=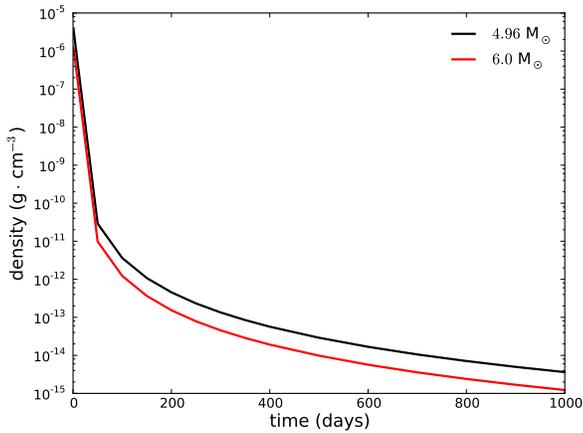,width=\columnwidth}
\caption{Density evolution for $4.96$ (black) and $6.0$ (red) M$_{\odot}$ enclosed mass subshells up to $1000$ days after the SN explosion.}
\label{fig:devolve}
\end{figure}

\begin{figure}
\psfig{file=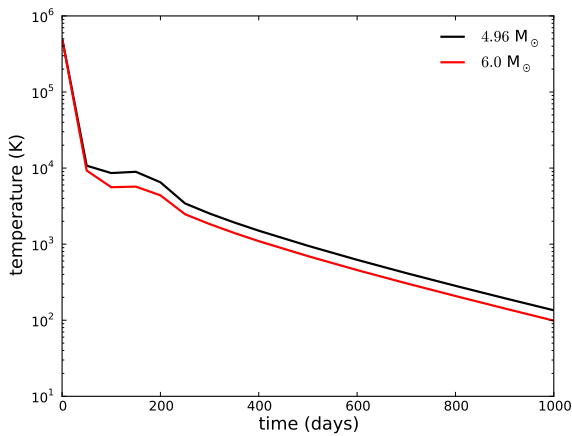,width=\columnwidth}
\caption{Temperature evolution for $4.96$ (black) and $6.0$ (red) M$_{\odot}$ enclosed mass subshells up to $1000$ days after the SN explosion.}
\label{fig:tevolve}
\end{figure}

Table \ref{tab:c} describes the data necessary for the calculation of
carbon grain formation. To compute the supersaturation of the
expanding gas, we find the phase equilibrium pressure by: $p_e=p_0
e^{-A/T+B}$, where $p_0$ is the standard pressure, $T$ is the temperature, and the values of
$A$ and $B$, listed here in Table \ref{tab:c}, are taken from Table 2
of \citet{NK03}. As the material ejected by the core-collapse supernova expands it also cools. Figures \ref{fig:devolve} and \ref{fig:tevolve} show the evolution of the density and temperature, respectively, of the expanding and cooling material for two of the enclosed mass subshells that we refer to in the rest of this work. 

\begin{table}
\centering
\caption{Carbon properties}
\begin{tabular}{ccccc}
        \hline
        \hline
        $A/10^4$ & & $\sigma$ & $v_{0}$ & $m_{0}$ \\
	(K) & $B$ & (erg$\cdot$cm$^{-2}$) & ($10^{-24}$ cm$^{-3}$) & ($10^{-23}$ g)\\ \hline
	$8.64726$ & $19.0422$ & $1400$ & $8.805$ & $1.995$\\
\end{tabular}
\label{tab:c}
\end{table}

We study four different values of the unknown sticking coefficient,
$\gamma = 1.0$, $0.1$, $0.01$, and $0.001$, neglecting any dependence
of $\gamma$ on the temperature of the gas and the size of the
cluster. For each value of the sticking coefficient, we study six
different values for the shape factor, $c = (36\pi)^{1/3}$, $5.4$,
$6.0$, $7.0$, $9.0$, and $12.0$, corresponding to shapes ranging from
a sphere to a flattened cylinder similar to a coin. We therefore
performed 24 simulations in total.  A sticking coefficient of
$\gamma=1.0$, and a shape factor for a sphere of $c = (36\pi)^{1/3}$ are
the usual parameters used in previous nucleation works.

\begin{figure}
\psfig{file=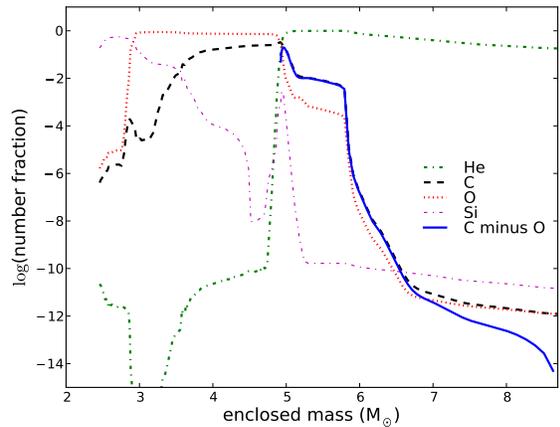,width=\columnwidth}
\caption{Number fraction of helium (green dash-dot), carbon (black dashed), oxygen (red dotted), and silicon (magenta dash-dot) atoms for enclosed
  masses from $2.45$ to $8.7$~M$_{\odot}$. The solid (blue) line is
  the carbon number fraction after the formation of CO molecules.}
\label{fig:cabun}
\end{figure}

As in \citet{NK03}, we assume that the stable formation of CO
molecules occurs prior to grain nucleation and that carbon grains will
form only where the initial number fraction of carbon is higher than
that of oxygen. Under this assumption, the number fraction of free carbon atoms available for dust formation is obtained simply from the initial number fraction of carbon minus the number fraction oxygen. We divide the expanding gases into a series of
enclosed mass shells beginning at $\sim 4.93$ M$_{\odot}$ and ending
at $\sim 6.21$ M$_{\odot}$. In this range of enclosed masses the
number fraction of carbon atoms, after the formation of CO, is
highest; ranging between $2\times 10^{-1}$ and
$8\times10^{-9}$. Figure \ref{fig:cabun} shows the number fraction of
carbon and oxygen atoms in the expanding ejecta from the hydrodynamic
results of \citet{UN02}. The solid (blue) line indicates the number
fraction of carbon left over after the formation of CO molecules. It
should be noted for completeness that grain nucleation can occur at
enclosed masses larger than we consider here, but is extremely
inefficient due to low carbon number fractions, $\sim10^{-14}$ and
below.

For each mass subshell, our code starts by following the evolution of
the density and temperature (see Figures \ref{fig:devolve} and \ref{fig:tevolve}, respectively, for examples) of the gas until the condition of
supersaturation is satisfied. From that point on, at each time step
the code performs three operations.
\begin{enumerate}
\item First, the code computes the number of critical clusters that are
  formed given the saturation, temperature, and partial pressure of
  the carbon atoms (according to Eq. (\ref{eq:Js})).
\item Second, the code grows any pre-existing grain formed at earlier
  times according Eq. (\ref{eq:dvdt}).
\item Finally, the code subtracts from the carbon in the gas phase the
  amount of carbon that has been locked in the solid phase by the processes in steps (i) and (ii). The concentration of
  gas-phase carbon is evolved according to:
  \begin{eqnarray*}
C_{1}(t_{n}) & = & C_{1}(t_{n-1}) - \frac{\Delta V_{grains,
    n}}{v_{0}V_{shell,n}}\\ 
&  &- \frac{C_{1}(t_{n-1})}{\tilde C_{1}(t_{n})}
\left(\tilde C_{1}(t_{n-1})-\tilde C_{1}(t_{n})\right),
\end{eqnarray*}
where $\Delta V_{grains, n}$ is the total change in volume of grains from the previous time step to the current time step. Then, the first term on the right-hand side is the concentration of
gas-phase carbon from the previous time step, the second term is the
change in the concentration of solid-phase carbon (it is a summation
of all grain changes, including formation of new critical clusters and the growth
of existing grains), and the final term accounts for the decrease in
concentration of gas-phase carbon due to the expansion of the shell.
\end{enumerate}
This process is repeated for each mass shell until the concentration
of gas-phase carbon is reduced to $1$ per cent of its original value. The
dust distributions from all shells are then summed together to produce
the final dust yield of each particular set $(\gamma,c)$.

\section{Results}

\subsection{Spherical Grains}

\begin{figure}
\psfig{file=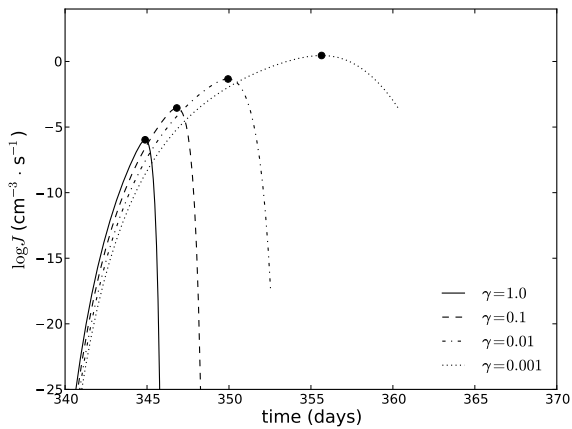,width=\columnwidth}
\caption{Nucleation rates for spherical grains ( $c=(36\pi)^{1/3}$),
  for the four considered sticking coefficients of $\gamma = 1.0$, $0.1$, $0.01$, and $0.001$. Filled circles
  indicate the time and rate of maximum nucleation. Rates are
  calculated at an enclosed mass coordinate $4.96M_{\odot}$.}
\label{fig:nrates4sc}
\end{figure}

After the supernova explosion, the hot gases expand and cool. The
cooling allows the gases to reach supersaturation conditions. Once the
supersaturation level becomes greater than unity, nucleation can
occur. The supersaturation level continues to rise and the nucleation
rate increases over time until depletion of available material becomes
significant and the supersaturation level begins to drop, after which
the nucleation rate falls off quickly. Eventually the gas is no longer
supersaturated and nucleation ceases. However, grain growth is still
possible.

Figure~\ref{fig:nrates4sc} shows the nucleation rate for spherically
shaped grains at enclosed mass coordinate $4.96$~M$_\odot$ for the
four sticking coefficients. We chose this particular enclosed mass
shell because it contains the highest abundance of carbon atoms, after
CO formation, of all our mass shells. The solid curve corresponds to
$c=(36\pi)^{1/3}$ and $\gamma=1.0$, the parameters generally used for
nucleation studies. The dashed, dash-dot, and dotted curves correspond
to $\gamma=0.1$, $0.01$, and $0.001$, respectively. To be consistent
with the works of \citet{KH87, KH89, KH91, NK03}, we consider the time
at which the nucleation rate is at its maximum to be the condensation
time of the grains. We show these condensation times as filled circles
at the maxima of the nucleation rates in the figure.

At early times, the reduced sticking coefficient makes the formation
of critically sized grains more difficult and results in a suppressed
nucleation rate. In the absence of strong nucleation, carbon atoms are
not depleted from the gas and the saturation continues to
increase. The reduced sticking coefficient thus causes the time at
which nucleation is at its maximum to be delayed, and the nucleation
to take place at higher saturation levels. As a consequence, a larger
number of critical clusters can form with much smaller size
(Eq. \ref{eq:nstar}).

\begin{figure}
\psfig{file=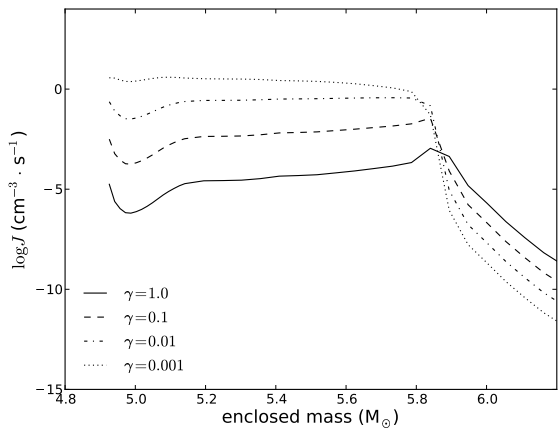,width=\columnwidth}
\caption{Maximum nucleation rates for spherical carbon grains at enclosed
  masses $< 6.2$ M$_{\odot}$ for four sticking coefficients.}
\label{fig:Js4sc}
\end{figure}

\begin{figure}
\psfig{file=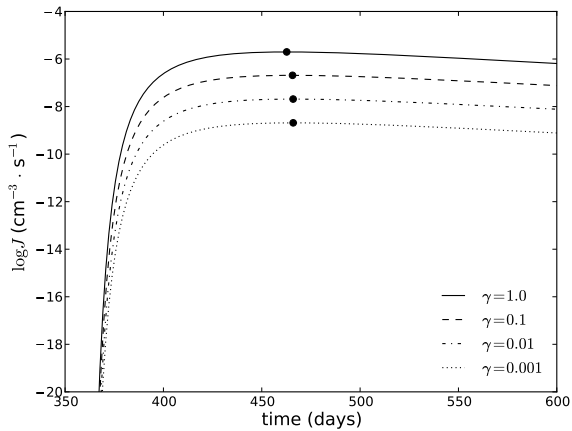,width=\columnwidth}
\caption{Nucleation rates for spherical carbon grains at an enclosed coordinate of $6.00$~M$_{\odot}$ for four sticking coefficients. Condensation times are indicated by filled circles.}
\label{fig:nrates60}
\end{figure}

Figure \ref{fig:Js4sc} shows the maximum nucleation rates for
spherical grains as a function of enclosed mass for our four sticking
coefficients. For enclosed masses up to M$_{r}\sim5.87$~M$_{\odot}$,
the nucleation rate maxima for reduced sticking coefficients exhibit
similar behaviour as in Figure \ref{fig:nrates4sc}. At greater enclosed
masses, however, the behaviour is inverted and a reduced sticking
coefficient results in depressed maximum nucleation rates. This
difference in behaviour is due to the reduction in the number fraction
of carbon available, from $\sim10^{-2.5}$ to $\sim10^{-5}$ (see Figure
\ref{fig:cabun}), which corresponds to a reduction in the
concentration of carbon monomers. Since the nucleation rate
(Eq.~(\ref{eq:Js})) is proportional to the concentration of monomers
squared, the drop in the carbon concentration consequently drops the
nucleation rate. Thus, the nucleation rates do not peak as strongly
(Figure \ref{fig:nrates60}), drawing out nucleation to later times for
all sticking coefficients, so that a catastrophic reduction in the
available material does not occur. Thus, the nucleation rates for
reduced sticking coefficients remain depressed throughout the
simulation time for higher enclosed mass shells.

\begin{figure}
\psfig{file=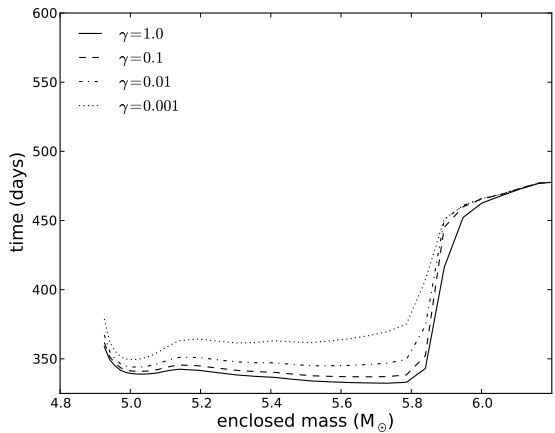,width=\columnwidth}
\caption{Condensation times of spherical carbon grain as a function of enclosed
  mass for four sticking coefficients.}
\label{fig:ctimes4sc}
\end{figure}

Figure \ref{fig:ctimes4sc} shows the condensation times corresponding
to the maximum nucleation rates shown in Figure \ref{fig:Js4sc}. The
condensation times of the solid curve ($c = (36\pi)^{1/3}$, $\gamma =
1.0$) are in good agreement with the condensation times for carbon
grains reported in \citet{NK03}. Here the lower sticking coefficients
result in delayed condensation times up to
M$_{r}\sim5.87$~M$_{\odot}$, outside which the condensation time is
only slightly delayed, or no longer delayed, when compared to the case
of $\gamma = 1.0$. The more noticeably delayed condensation times, as
well as the similarity of condensation times for all our sticking
coefficients, at higher enclosed mass shells are also due to the drawn
out nucleation process already discussed above.

\begin{figure}
\psfig{file=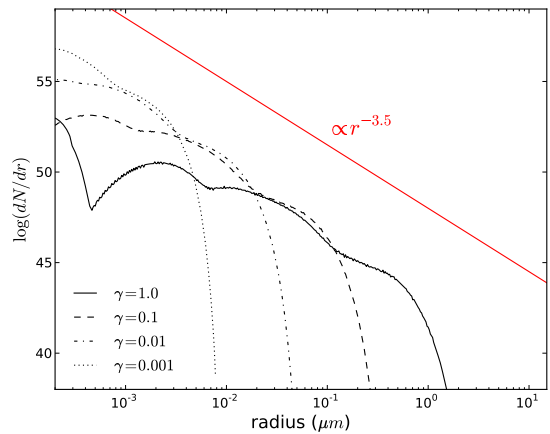,width=\columnwidth}
\caption{Size distribution of carbon grains for $\gamma = 1.0$, $0.1$,
  $0.01$, $0.001$, with $c = (36\pi)^{1/3}$. For reference, the solid (red) line
  represents the power-law distribution with the form of $N_{r}~\propto~r^{-3.5}$, which has been suggested as that of interstellar grains (e.g.,~\citet{MR77}).}
\label{fig:4scsizes}
\end{figure}

\begin{figure}
\psfig{file=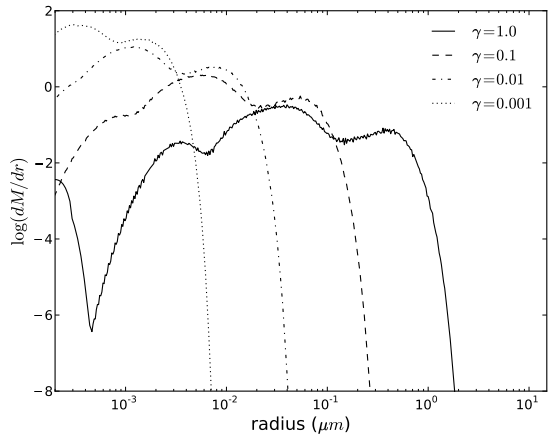,width=\columnwidth}
\caption{Mass distribution of carbon grains for $\gamma = 1.0$, $0.1$,
  $0.01$, $0.001$, with $c=(36\pi)^{1/3}$.}
\label{fig:4scmassdist}
\end{figure}

Reducing the sticking coefficient makes both nucleation and grain
growth more difficult. A reduced sticking coefficient results in a
larger number of smaller grains, fewer larger grains, and a smaller
maximum grain radius. In Figures \ref{fig:4scsizes} and
\ref{fig:4scmassdist}, respectively, we show the size and mass
distributions for spherical grains. For the case of $\gamma = 1.0$,
the maximum grain size achieved is between $2$ and $3~\mu$m. When the
sticking coefficient is reduced to $0.001$, the maximum grain size is
decreased to less than $0.01~\mu$m.

Since the reduced sticking coefficient causes larger numbers of small
grains to form, the small grains contain more relative mass than the
larger grains, as can be seein in Figure
\ref{fig:4scmassdist}. However, the total mass of the grains is
relatively robust. The total masses of dust grains for the spherical
case are shown in Table \ref{tab:nonspheremasses}, along with the
aspherical cases which are discussed in the next section. Even though
the onset of nucleation is delayed due to the reduced sticking
coefficient at enclosed masses less than $5.87$ M$_{\odot}$, where the
majority of available carbon is contained, the subsequent grain growth
consumes almost all of the carbon atoms for $\gamma > 0.001$. Thus,
the total mass of carbon dust is principally determined by the mass of
pre-existing carbon atoms.

\subsection{Non-spherical Grains}

\begin{figure}
\psfig{file=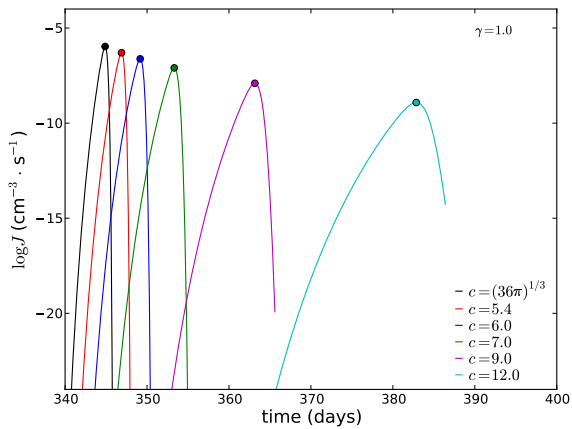,width=\columnwidth}
\caption{Nucleation rates as a function of time for six shape factors
  with $\gamma = 1.0$. Filled circles indicate maximum nucleation
  rate. Rates shown are for nucleation within a shell of enclosed mass of
  $\sim4.96$ to $\sim4.97$ M$_{\odot}$.}
\label{fig:nrates6sf}
\end{figure}

We also calculated nucleation rates for aspherical grains. We choose a
range of shape factors from $c = 5.4$ to $c = 12.0$. While $c=6.0$ is
the shape factor of a cube, each shape factor can correspond to a
number of different grain shapes. The shape factor can be thought of
as a deviation from the spherical case, the bigger the shape factor,
the larger the deviation from a sphere and the larger the surface for
a given volume.

\begin{figure*}
\psfig{file=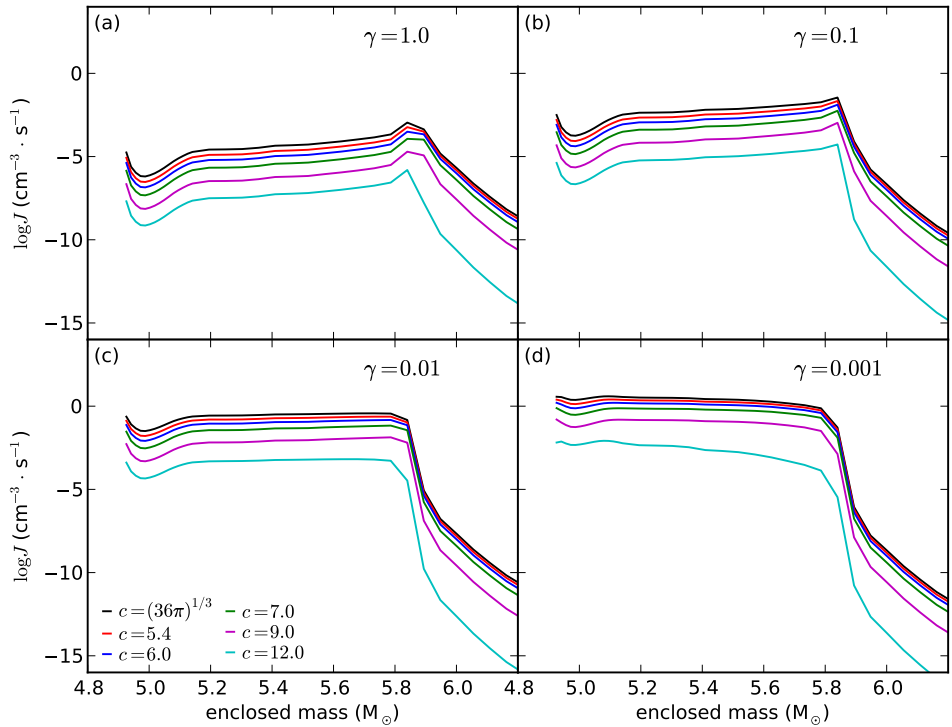, width=1.75\columnwidth}
\caption{Maximum nucleation rates for all six shape factors and four sticking
  coefficients.}
\label{fig:nrates6}
\end{figure*}

\begin{figure*}
\psfig{file=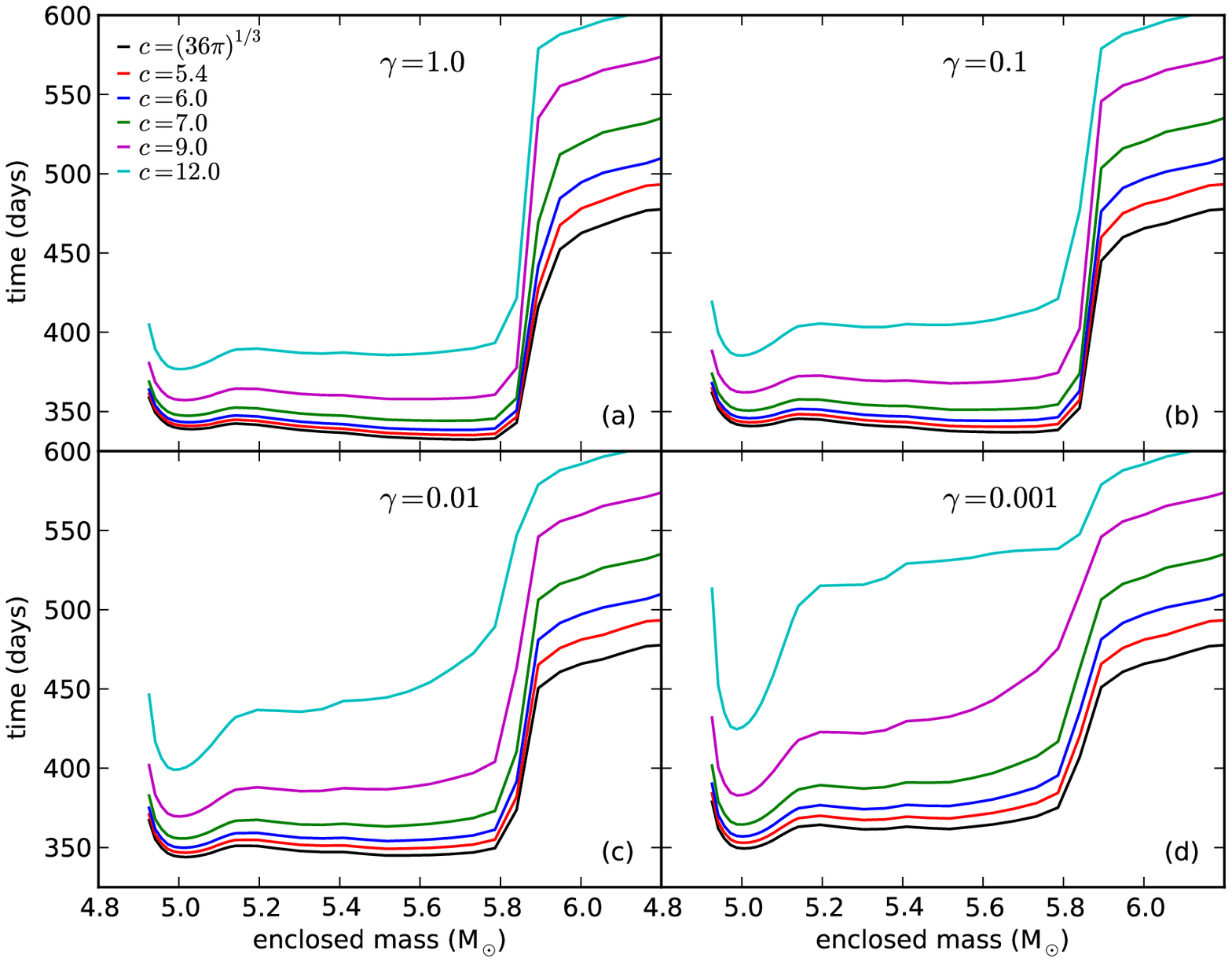, width=1.75\columnwidth}
\caption{Condensation times for six shape factors and four sticking
  coefficients.}
\label{fig:ctimes6}
\end{figure*}

In Figure \ref{fig:nrates6sf} we show the nucleation rates for the
same enclosed mass shell as shown in Figure \ref{fig:nrates4sc}, for
increasing shape factors with $\gamma = 1.0$. Since the shape factor
appears in the exponential term of the nucleation rate equation
as $J_{s}\propto \exp{(-c^{3})}$ (Eq. (\ref{eq:Js})), the higher shape
factors reduce the nucleation rate. Therefore, the higher
supersaturation levels at later times need to be attained so that the
nucleation rate becomes high enough for significant depletion of the
gas due to the growth of newly formed grains.  However, the maximum
nucleation rates for increasing shape factors are decreased. This is
due to the fact that with a larger surface to volume ratio aspherical
grains tend to grow faster, producing a sizeable depletion of the
gas-phase carbon even for moderate nucleation rates. For all mass
coordinates, increasing the shape factor leads to lower nucleation
rates and delayed condensation times (Figures \ref{fig:nrates6}a and
\ref{fig:ctimes6}a, respectively).

In Figure \ref{fig:nrates6}b--d we show the nucleation rate maxima for
all shape factors for each of the other three sticking
coefficients. For any sticking coefficient, the shape factor has the
same effect of suppressing the nucleation rate. All condensation
times, shown in Figure \ref{fig:ctimes6}b--d, increase with the
increased shape factor. Furthermore, for reduced sticking
coefficients, greater shape factors lead to greater delays in the
condensation time.

The shape factor $c=(36\pi)^{1/3}$ is the only shape factor that has
just one associated shape, the sphere. The other shape factors we
consider do not necessarily have a unique grain shape. To find a size
distribution that is easily comparable to the spherical case, we
define a {\it volume equivalent radius} for the aspherical grains as:
\begin{equation}
r_{\rm{eff}}=\left(\frac{3V}{4\pi}\right)^{1/3},
\end{equation}
where $V$ is the grain volume.

\begin{figure*}
\psfig{file=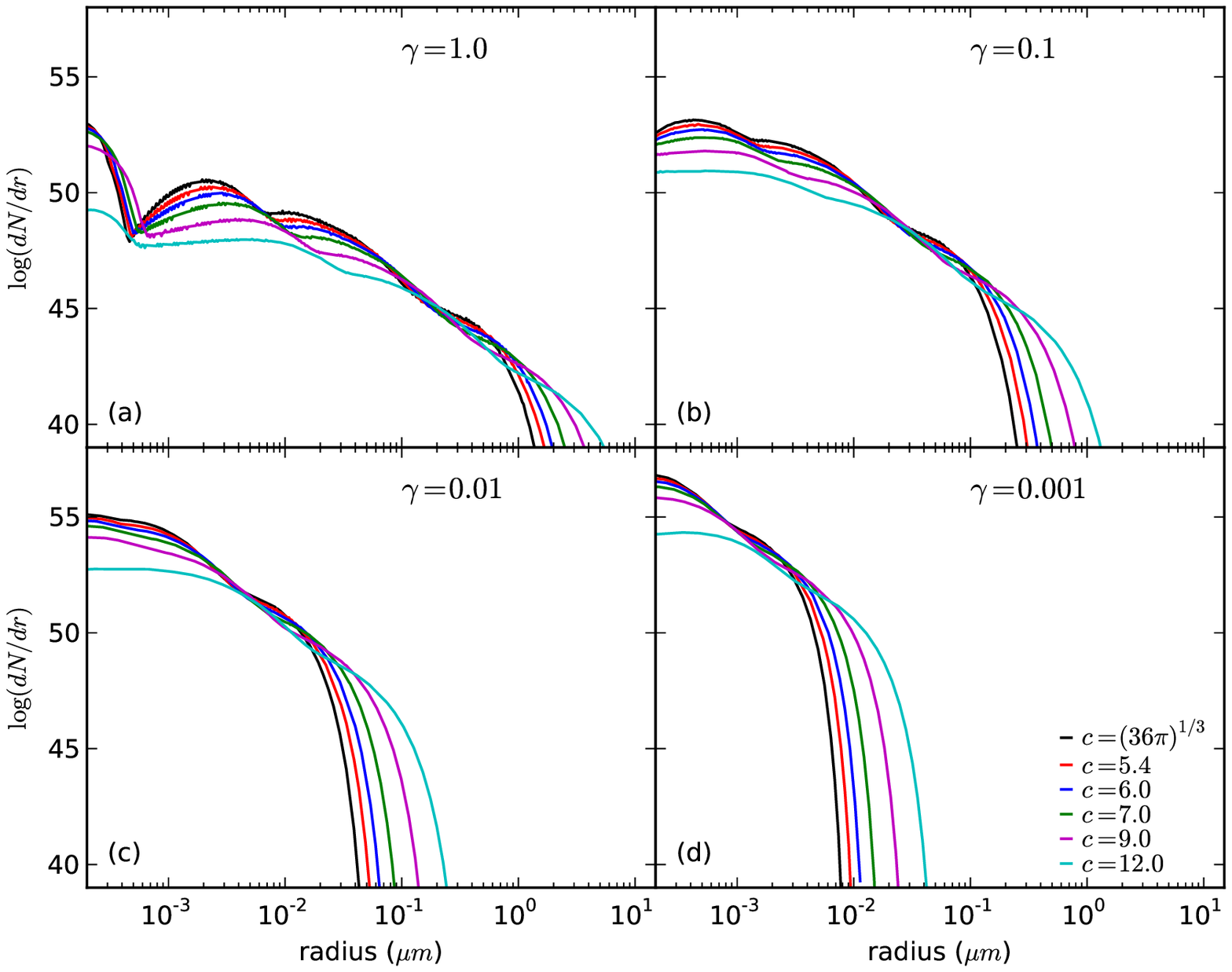, width=1.75\columnwidth}
\caption{Size distribution of carbon grains for six shape factors and
  four sticking coefficients.}
\label{fig:sizes6}
\end{figure*}

\begin{figure*}
\psfig{file=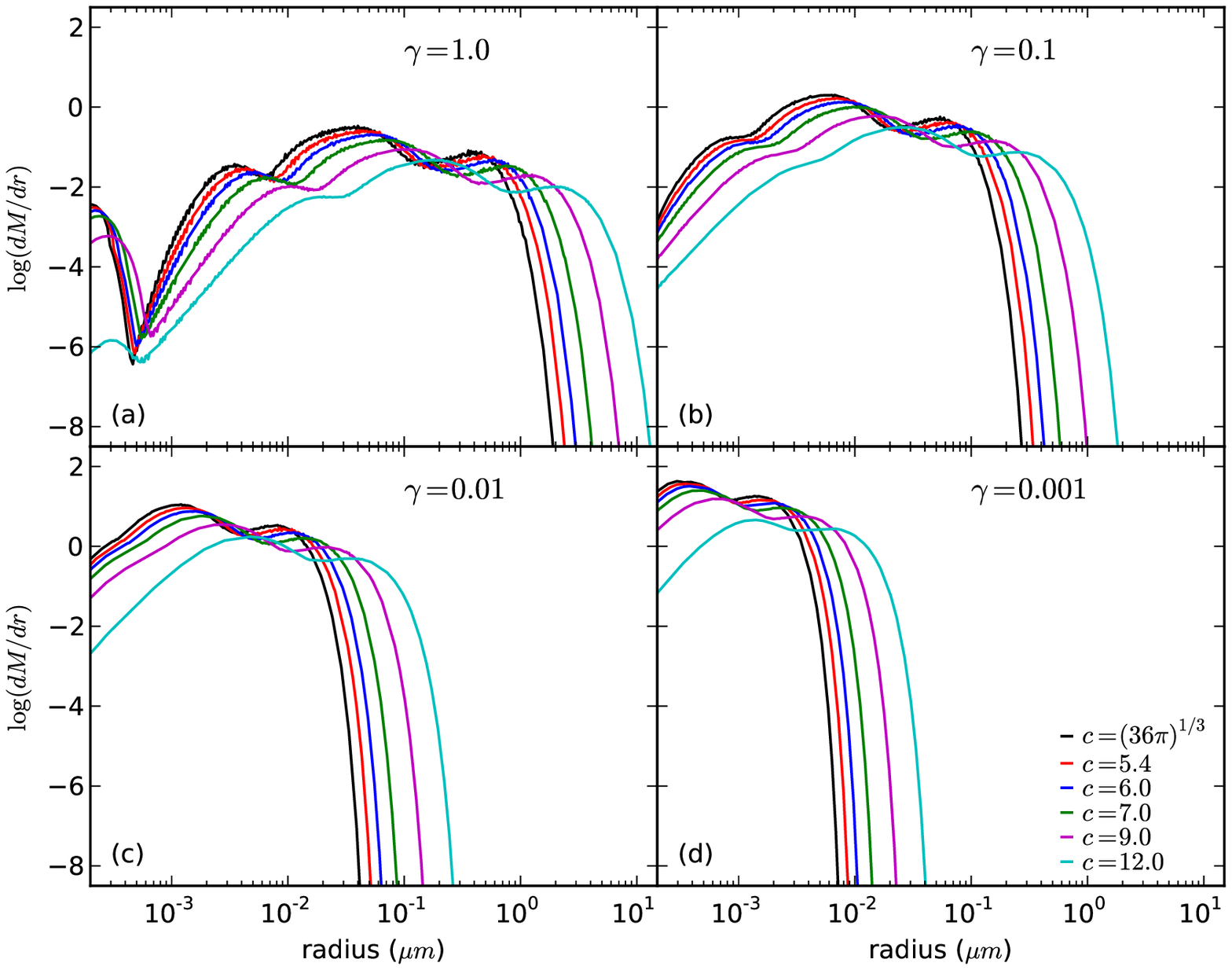,width=1.75\columnwidth}
\caption{Mass distribution of carbon grain for six shape factors and
  four sticking coefficients.}
\label{fig:masses6}
\end{figure*}

Figures \ref{fig:sizes6}a--d show the size distributions for all shape
factors and sticking coefficients. As the shape factor is increased,
the size and number of larger grains also increases. We find that the
largest maximum grain radii are formed with $c = 12.0$ and $\gamma =
1.0$, and have a volume equivalent radius of almost $18~\mu$m. On the other
hand, the smallest maximum grain radii ($\sim 0.008~\mu m$) are formed
with $c=(36\pi)^{1/3}$ and $\gamma = 0.001$. As the sticking
coefficient is reduced, the maximum grain radii for each shape factor
are also reduced, and the number of smaller grains is increased.  Even
though there is a much larger number of small grains than large
grains, the majority of the dust mass is contained within intermediate
sized grains. In Figure \ref{fig:masses6}a ($\gamma = 1.0$), most of
the dust mass is contained in grains with volume equivalent radii between
$0.01$ and $0.5~\mu$m.

\begin{figure}
\psfig{file=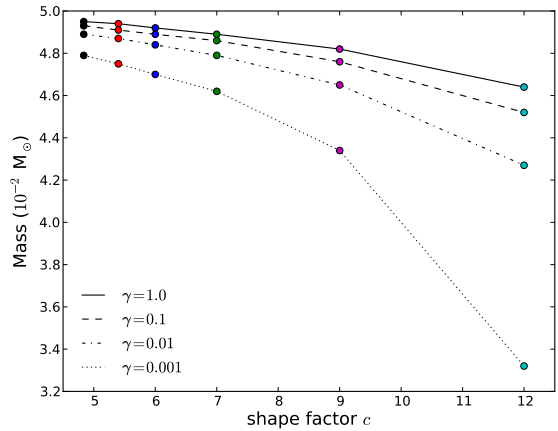,width=\columnwidth}
\caption{Total mass of carbon dust formed for all simulations.}
\label{fig:totalmass}
\end{figure}

The masses of carbon grains formed (Figures \ref{fig:masses6}a--d) are
dominated by the relatively small numbers of larger sized grains as
the shape factor is increased. However, as the sticking coefficient is
reduced (Figures \ref{fig:masses6}b--d), even the masses of the grains
formed with the largest shape factor become dominated by the smaller
sized grains. The total mass of carbon grains formed for increased
shape factors is given in Table \ref{tab:nonspheremasses} and shown in
Figure \ref{fig:totalmass}. Again, we notice that despite the
difference in the size distribution due to different microphysical
parameters, the total mass of dust that condenses is fairly robust
(within a factor of $1.5$) and constitutes almost the total amount of
carbon that was left in the gas phase after the creation of CO
molecules.

\begin{table}
\centering
\caption{Total mass of carbon grains formed}
\begin{tabular}{r|cccc}
	\hline
	\hline
	\multicolumn{5}{c}{Mass ($10^{-2}$~M$_{\odot}$)} \\
	\hline
	$c$ & $\gamma =1.0$ & $\gamma = 0.1$ & $\gamma =0.01$ & $\gamma =0.001$ \\
	\hline
	$(36\pi)^{1/3}$ & $4.95$ & $4.93$ & $4.89$ & $4.79$\\
	$5.4$   & $4.94$ & $4.91$ & $4.87$ & $4.75$\\
	$6.0$   & $4.92$ & $4.89$ & $4.84$ & $4.70$\\
	$7.0$   & $4.89$ & $4.86$ & $4.79$ & $4.62$\\
	$9.0$   & $4.82$ & $4.76$ & $4.65$ & $4.34$\\
	$12.0$ & $4.64$ & $4.52$ & $4.27$ & $3.32$\\
\end{tabular}
\label{tab:nonspheremasses}
\end{table}

\begin{figure*}
\psfig{file=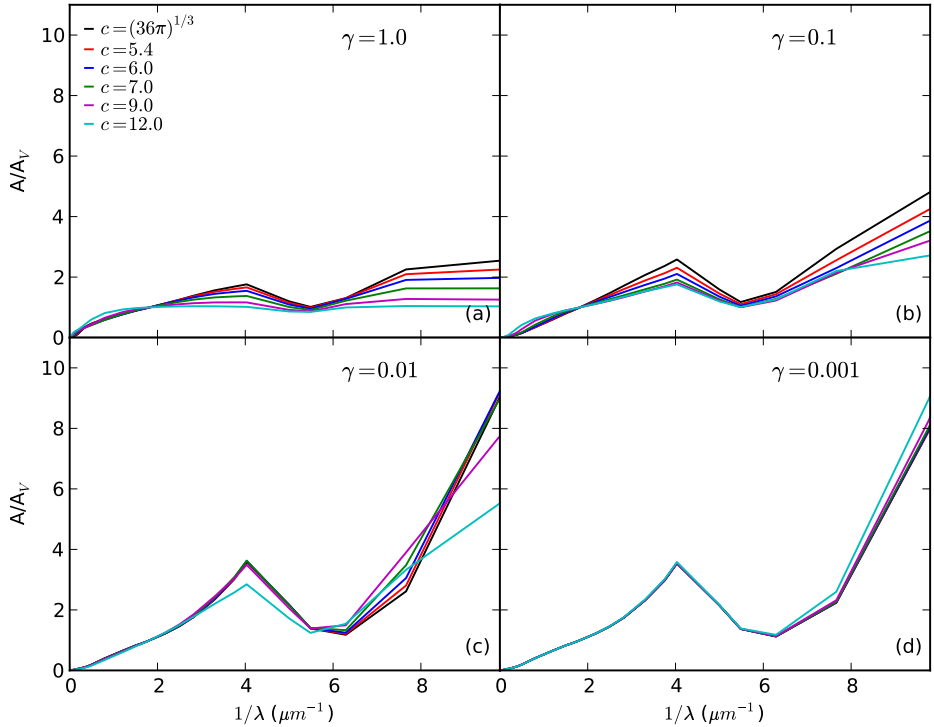,width=1.75\columnwidth}
\caption{Extinction curves for each sticking coefficient and shape factor.}
\label{fig:extinct}
\end{figure*}

\begin{figure}
\psfig{file=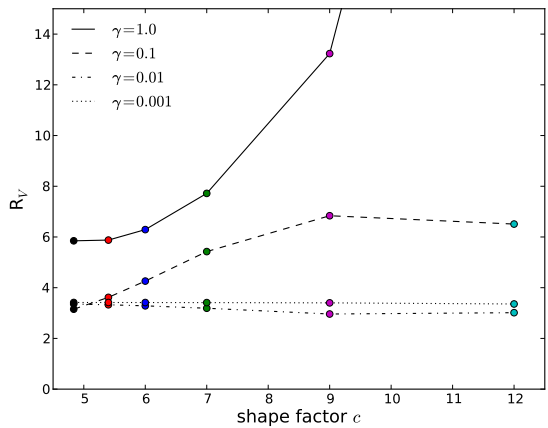,width=\columnwidth}
\caption{R$_V$ values for 23 simulations. Not shown is R$_V = 41$ for $c=12.0$ and $\gamma = 1.0$.}
\label{fig:Rv}
\end{figure}

\section{Discussion}

We have shown that varying the microphysical properties of dust grains
has important effects on the condensation times, nucleation rates, and
size distributions of carbon dust grains from type II supernova
explosions. However, the total mass of dust is only modestly affected
by the changes in the grain properties. An inadequate choice of the shape or
sticking coefficient is not therefore a viable explanation for the
discrepancy between the mass of dust grains predicted in SN explosions
and the observed dust mass in local type II SNe (see Section 1 for a
more thorough discussion and references).

We find that a larger saturation is necessary to achieve efficient
nucleation with either a small sticking coefficient ($\gamma<1$) or
for aspherical grains ($c>(36\pi)^{1/3}$). For that reason, all our
simulations show that the condensation time grows when sticking coefficients less than unity or grain shapes that are different from spherical are adopted. However, differences can be found in
the nucleation rates and final size distribution of the
grains. Simulations with a low value of the sticking coefficient show
a delayed nucleation but very high nucleation rates, thereby producing
large quantities of small grains. This is due to the fact that low
sticking coefficients inhibit both nucleation and grain growth and,
therefore, all the carbon remains in the gas phase until a high
saturation level is reached. At that point, many small grains are
nucleated and the atomic carbon is quickly depleted. \citet{BS07} found similar effects when calculating dust nucleation with a sticking
coefficient of $\gamma = 0.1$. Asphericity of the grains, on the other
hand, inhibits nucleation but enhances grain growth. As a consequence,
even if fewer grains are nucleated, they grow fast and the result is a
grain size distribution characterized by less numerous, larger grains.

\begin{figure}
\psfig{file=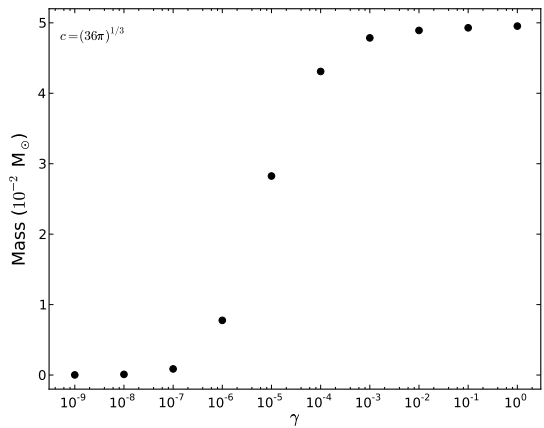,width=\columnwidth}
\caption{Total mass of dust formed for spherical carbonaceous grains with sticking coefficients down to $\gamma = 10^{-9}$.}
\label{fig:lowgamma}
\end{figure}

We find that the total mass of carbonaceous dust formed remains relatively stable even with sticking coefficients as low as $0.001$. For the spherical case only, we explored the possibility of even smaller sticking coefficients, down to $\gamma = 10^{-9}$. We find that the mass of carbon dust formed becomes significantly reduced for sticking coefficients of $\gamma = 10^{-7}$ and below (see Figure \ref{fig:lowgamma}). With sufficiently low values of sticking coefficient (below $10^{-8}$) there is virtually no dust formation, but the required sticking coefficients seem unphysically low. 

In terms of the observable properties of the SN-condensed dust, we
find that the quantity that is most affected is the extinction curve
(see Figure~\ref{fig:extinct}). Not surprisingly, simulations with
small sticking coefficient (which, as explained above, produce large
amounts of small grains), result in a very steep extinction curve at far-UV wavelengths,
with R$_V$ values between 3 and 3.5, shown in Figure \ref{fig:Rv}. We make special note that the R$_V$ values for $\gamma =1.0$ are relatively high, even though the size distribution of the grains is consistent with that of interstellar grains (see Figure \ref{fig:4scsizes}), because we account for only carbon grains and that including other grains, such as silicates, could decrease the R$_V$ values. On
the other hand, simulations with very aspherical grains and relatively
high sticking coefficients produce larger grains and, consequently,
grey extinction curves. Without the knowledge of the values of the
sticking coefficient and of the shape factor it is therefore
impossible to predict the extinction curve of SN-condensed dust. This
is a particularly worrying conclusion since the extinction curve is
relatively easy to measure, even at high redshift, and could be used
as an observational constraint for the origin of dust in the various
environments.  For example, \citet{MS04} compared the extinction curve
measured in a quasar at $z=6.2$ to the extinction curve calculated using the dust model by
\citet{TF01}. They find that the data and the theoretical prediction
are in good agreement and conclude that the dust observed in
SDSSJ104845.05+463718.3 is indeed condensed in SN explosions (see also
\citet{STR07}). In light of our results, such conclusions need
confirmation once a complete theory of SN dust nucleation is obtained.

The dust that condenses, however, is not the dust that is ejected into
the ISM. Dust produced in a CCSN has to travel through the reverse
shock before being released into the ISM. The reverse shock can
destroy most of the dust, in particular the smaller dust grains
\citep{DS79, NK07, NL08, SS10}. If astrophysical dust formation is indeed
characterized by small values of the sticking coefficient, it is
likely that the amount of dust formed is reduced significantly by the
reverse shock. Conversely, increased shape factors allow for the
formation of larger grains which would survive shock processing. The
microphysical properties of dust grains can therefore affect the mass
of dust that is injected in the interstellar medium, even
though they affect only marginally the dust that is condensed during
the early stages of the explosion.

Grain nucleation with non-spherical shapes may be more complicated than
we considered here. We have assumed that the shape factors of the
grains do not change as the grains grow. Since we assume here that grains grow by
the addition of monomers, the shape of small clusters can change as
monomers attach, in turn altering the shape factor. Another route to
take could be to nucleate grains at an arbitrary shape factor and then
allow the grains to grow into spheres. For example, one may turn
grains into spheres when the number of monomers is larger than a given
value.  In this case, increased shape factors may not lead to such
large grains as we show in our size distributions (Figure
\ref{fig:sizes6}).

This work is purely a parametric study that aims to show that the
microphysical properties of the grains are important, but not to point
to any specific values of $\gamma$ and $c$ to be used in nucleation
calculations. Therefore, we have chosen to neglect important factors that
need be considered in a complete SN dust nucleation model. These include the
choice of progenitor model, the presence of other dust species (see \citet{NK03}) as well as charged molecules that may interfere with carbonaceous dust condensation, and the destruction of CO molecules, due to photodissociation or collisons with fast electrons and charged particles \citep{PD+89, LDM90, LDL92, CDM01}, that can inject additional carbon atoms into the available
monomer concentration (see \citet{TF01, BS07}).

A zero-metallicity 20 M$_{\odot}$ CCSN progenitor model was chosen because SNe at high redshift are expected to have zero metallicity. In general, the relative abundances of major elements in the He core are not significantly different among the SN models with different metallicities. Thus, the species of dust formed do not depend on the metallicity of the SN progenitor star \citep{NK10}. Additionally, the gas density and temperature in the He core are almost independent of the progenitor mass and metallicity as long as the kinetic energy of the explosion is the same \citep{NK03}. Therefore, SN models with non-zero metallicities, or with different progenitor masses, are expected to show similar effects on carbon grains formation as we see here. This means that the mass of carbon dust formed in the SN ejecta is rather insensitive to changes in the sticking coefficient and shape factor and is purely determined by the mass of carbon atoms available for dust formation in the He layer. However, to confirm such expectations, additional progenitor models would need to be investigated. 

Perhaps more important than the choice in progenitor model is the
dissociation process of CO molecules. In this paper we assumed the
formation of CO molecules to be complete and considered only the
condensation process of C grains in the He layer where C/O $> 1$.
In the expanding ejecta, CO molecules could be destroyed through
interactions with fast electrons from radioactively decaying $^{56}$Co
and charged particles such as He$^{+}$ and Ne$^{+}$ \citep{PD+89, LDM90, LDL92, CLD99, KCF00, CDM01, DCM06}, allowing for more free carbon (and oxygen)
atoms to be available for grain formation than we consider here.
However, the number abundance of silicon atoms is too small for
most of the enclosed mass regions (M$ = 4.93$--$6.21$ M$_{\odot}$) in this
work (see Figure \ref{fig:cabun}), so that the formation of SiC and silicate gains
cannot be expected. Therefore, the dissociation of CO molecules
due to interactions with He$^{+}$ only results in a slight enhancement of
the final mass of carbon grains and never changes our conclusion on
the dependence of formation process of C grains on the microphysical
properties.

On the other hand, \citet{CLD99} and \citet{DCM06}
show that CO dissociation enables carbon dust grains to form even
in O-rich layer where C/O $< 1$. Given that the abundance of silicon
atoms in the O-rich layer is higher than in the He layer, the formation of SiC
grains could be expected there. However, as discussed in \citet{NK03}, even if free carbon and silicon atoms coexist abundantly,
the nucleation theory does not predict the formation of SiC grains.
The formation process of large SiC grains as appeared in presolar
grains, as well as formation efficiency of molecules is to be pursued
in more sophisticated studies of dust formation. Furthermore, in the
O-rich layer, the formation of silicate grains is also feasible. \citet{BS07} show that the formation of silicate grains is more
sensitive to changes in sticking coefficient than carbon grains, and the
mass of silicate grains formed can be reduced for even $\gamma = 0.1$.
The inclusion of silicate grains could affect the resulting extinction
curves.

We have adopted the thermodynamic approach for this study because it
involves the simplest nucleation equations, however, use of the
kinetic theory of nucleation should be considered in the future. The
kinetic theory still needs to take the sticking coefficient into
account, but makes consideration of an evolving shape factor
unnecessary, because the shape of the grain from a complex solid (for
a few molecules) to a sphere (for $\sim 100$~molecules) is
intrinsically taken into account.

\section*{Acknowledgements}

We would like to thank the reviewer for their insightful comments and suggestions for the improvement of this work. We would also like to thank Raffaella Schneider and Andrea Ferrara for their comments. Thanks go to the Institute for the Physics and Mathematics of the
Universe (IPMU), University of Tokyo, Kashiwa, Japan, for their
generous hospitality, and to the NSF East Asia and Pacific Summer
Institute (EAPSI) Japan 2010 program (award \#1015575) and the
Japanese Society for Promotion of Science for their generous support. This work has also been supported in part by World Premier International Research Center Initiative, MEXT, Japan.

\end{document}